\documentclass[letterpaper,showpacs,showkeys,onecolumn,eqsecnum,superscriptaddress,floatfix]{revtex4}

\usepackage{amssymb}
\usepackage{amsmath}
\usepackage{latexsym}
\usepackage{epsfig}
\usepackage{graphicx}
\usepackage{color}
\usepackage{array}
\usepackage[linktocpage]{hyperref}
\usepackage{epstopdf}
\usepackage[english]{babel}

\begin{document}

%
%
%
\title{An independent search of gravitational waves in the first observation run of advanced LIGO using cross-correlation} 
\author{Javier M. Antelis} 
\email[]{mauricio.antelis@itesm.mx}
\affiliation{Tecnologico de Monterrey, Escuela de Ingenier\'ia y Ciencias \\
Av. Gral. Ram\'on Corona 2514, Zapopan, Jal., 45201, M\'exico}
\author{Claudia Moreno} 
\email[]{claudia.moreno@cucei.udg.mx} 
\affiliation{Departamento de F\'isica,
Centro Universitario de Ciencias Exactas e Ingenier\'ias, Universidad de Guadalajara\\
Av. Revoluci\'on 1500, Colonia Ol\'impica C.P. 44430, Guadalajara, Jalisco, M\'exico}

%

%
\bigskip
\begin{abstract}
This work describes a template-free method to search gravitational waves (GW) using data from the LIGO observatories simultaneously.
The basic idea of this method is that a GW signal is present in a short-duration data segment if the maximum correlation-coefficient between the strain signals is higher than a significant threshold and its time difference is lower than the 10 ms of inter-observatory light propagation time. 
Hence, this method can be used to carry out blind searches of any types of GW irrespective of the waveform and of the source type and sky location.
An independent search of injected and real GW signals from compact binary coalescences (CBC) contained in the first observation run (O1) of advanced LIGO was carried out to asses its performance.
On the basis of the results, the proposed method was able to detect GW produced by binary systems without making any assumption about them.
\end{abstract}

%
\pacs{
04.30.-w,  
95.30.Sf,  
07.05.Kf   
}

\keywords{
Gravitational Waves,  %
LIGO,
Detection, 
Correlation,
Time-Lag
}

%
\maketitle


\section{Introduction}
\label{sec:introduction}
%
%
%
The LIGO-VIRGO scientific collaboration has reported six gravitational waves (GW) events, which have offered a new window to explore and study the universe.
The events GW150914, GW151226, GW170104, and GW170608 were the first observations of GW produced from merging stellar-mass binary black holes (BBH) and were detected by the LIGO Hanford and Livingston twin observatories with a difference of time lower than the 10 ms of inter-observatory light travel time \cite{PhysRevLett.116.061102,PhysRevLett.116.241103,PhysRevLett.118.221101,2041-8205-851-2-L35}.
The event GW170814 was produced from a binary black hole coalescence and it also was detected by the VIRGO observatory making it the first three-observatory detection of GW, which enhanced the capability to localize the source on the sky with more detail \cite{PhysRevLett.119.141101}.
The last event, GW170817, was the first observation of GW produced from a colliding binary neutron star (BNS) which allowed to follow up its transient counterparts across the electromagnetic spectrum \cite{PhysRevLett.119.161101}. This first joint observation of GW and electromagnetic radiation opened a new world of possibilities to understand known and unexplored astrophysical objects.
These groundbreaking astrophysical observations have not only confirmed Einstein's general relativity theory presented one century ago \cite{Einstein15a,Einstein15b}, but have also revealed the existence of exotic astrophysical objects as binary black holes and binary neutron stars coalescences.
This new GW-based astronomy in conjunction with the well-established electromagnetic-based astronomy has marked the beginning of the era of the multi-messenger astrophysics \cite{2041-8205-848-2-L12,2041-8205-848-2-L13}.

%
The detection of GW was the result of long-term theoretical research devoted understand their fundamental physics \cite{Maggiore2007,Carroll04a,Shapiro} and experimental research carried out to develop a network of ultra sensible instruments that are able to measure extremely tiny strain signals induced by GW from astrophysical origin \cite{0034-4885-72-7-076901,0264-9381-32-7-074001,AdvancedVirgo2015}.
Another essential component that enabled the observation of GW is the data analysis and processing methods whose aim is to detect the GW signals and to characterize and localize its source from the noisy measurements obtained from the observatories.
Indeed, the observation of the six GW events involved the use of the matched filter algorithm \cite{PhysRevD.85.122006,PhysRevD.71.062001} and thousands of compact binary coalescence candidate waveforms or templates computed with the effective-one-body formalism \cite{PhysRevD.62.064015} which combines post-Newtonian solutions \cite{poisson2014gravity,1742-6596-1030-1-012005} with results from perturbation theory \cite{Moreno:2016urq} and numerical relativity \cite{Baumgarte2010}.
The matched filter is the optimal method to search signals embedded in highly noisy observations and consists in the frequency-domain correlation between the observed data and a template.
Hence, many templates are required to cover the multidimensional parameter space of the source which includes masses, spins, among others.
The above leads to the following drawbacks, first, the computational burden due to the extremely high number of required templates, second, the possibility of missing detections due to no template matches with the real GW signal, and third, it is not possible to detect GW signals whose waveforms are unknown in advance as burst GW signals emitted by core-collapse supernovae \cite{Gill:2018hxg,Klimenko:2008fu}.
%
%
%

%
As an independent alternative to carry out a straightforward search of GW, this work presents a fundamental computational tool to detect GW that does not assume waveform models and employs the data from the detectors simultaneously.
The method computes the cross-correlation between short-duration cleaned strain data from the LIGO observatories and determines the existence of a GW signal provided that the maximum correlation-coefficient is greater than a significant threshold and its corresponding time-lag is between the inter-observatory light travel time (i.e., $\pm 10$ ms for the LIGO Hanford and Livingston observatories as their separation is 3 Km or equivalently 10 ms at the speed of light).
Hence, transients that coincide in the two detectors and that have a phenomenological meaning are automatically detected while no assumptions about the signature of the GW signals are needed, i.e., template-free search method.
%
%
%
Because of this, the method can be used to search any GW regardless of its waveform and therefore  of its astrophysical source and location.
The method was applied to search injected and real GW signals contained in LIGO's first observation run (O1) and the detection accuracy results showed that it is was possible to detect most of the CBC hardware injection and the first GW ever detected, the event GW150914.
Previous studies have employed cross-correlation to understand the first GW event \cite{1475-7516-2016-08-029}, nonetheless, the present work shows a more detailed technical cross-correlation based method that automatically generates detection triggers along with its application in an exhaustive search using LIGO data from O1.
The manuscript will organize as follows. Section II describes the technical details of the method and its implementation. Section III presents the results of the application of the method to search injected and real GW signal using LIGO data from O1. Section IV discusses the results and the properties of the method. Finally, Section V presents the conclusion.

\section{Description of the method}
\label{sec:method}

\subsection{Problem statement}
Let $s_{H}^{i}(t)=[s_{H}^{i}(t_{0}), s_{H}^{i}(t_{1}), \dots ,s_{H}^{i}(t_{N-1})]^{T}$ and $s_{L}^{i}(t)=[s_{L}^{i}(t_{0}), s_{L}^{i}(t_{1}), \dots ,s_{L}^{i}(t_{N-1})]^{T}$ be two uniformly distributed discrete time series in an $i$-$th$ slice of cleaned strain data from the LIGO Hanford and Livingston observatories.
The sampling period is $t_{s} = t_{j}-t_{j-1}$ and the sampling frequency is $f_{s} = 1/t_{s}$, thus, the duration is $T_{slice}=N/f_{s}$ seconds where $N$ is the total number of samples.
The duration of this slice ranges from milliseconds up to a few seconds to encompass observable GW signals in the sensitivity frequency band of LIGO.
%
%
%
These two signals can be modeled as:
\begin{eqnarray}
s_{H}^{i}(t) &=& n_{H}(t) + h(t)
\\
s_{L}^{i}(t) &=& n_{L}(t) + (-1)^{l} h(t+\tau)
\end{eqnarray}
where $n_{H}(t)$ and $n_{L}(t)$ are the noise in the two detectors (which are uncorrelated and independent), $h(t)$ is an unknown GW signal (whose duration, arrival time and waveform are not known), $\tau$ is the time lag introduced by the different arrival time of the GW signal to both detectors ($-10<\tau<10$ ms as the separation between the detectors is 3 Km, which is equivalent to about 10 ms at the speed of light) and $l \in \{0,1\}$ accounts for the possible inversion between the GW signal received in the two interferometers (originated by the direction of the traveling GW).

%
The goal of the GW detection problem is to determine if the two signals in the slice of cleaned strain data are only noise (i.e., no GW signal is present) or if they consist of noise plus an unknown GW signal (i.e., a GW signal is present).
To address this problem, this work proposes a method to generate a slice marker that indicates whether an unknown GW signal is absent or present and to automatically estimate the time-shift, the cross-correlation value, and the signal inversion factor if a GW is detected.

\subsection{Detection method}
The proposed detection method computes the cross-correlation to measure the similarity across time-shifts between the two signals from the LIGO Hanford and Livingston observatories and decides whether a GW signal is present based on the phenomenological information.
%
%
The idea is that if the maximum correlation-coefficient between the two signals is superior to a significant threshold (which is computed directly from the data) and its corresponding time-shift is between $-10$ and $10$ ms (i.e., the maximum inter-observatory propagation time), then a GW signal is present, otherwise, a GW is absent.

\begin{figure*}[t]
	\begin{center}
  	\includegraphics[width=1.00\textwidth]{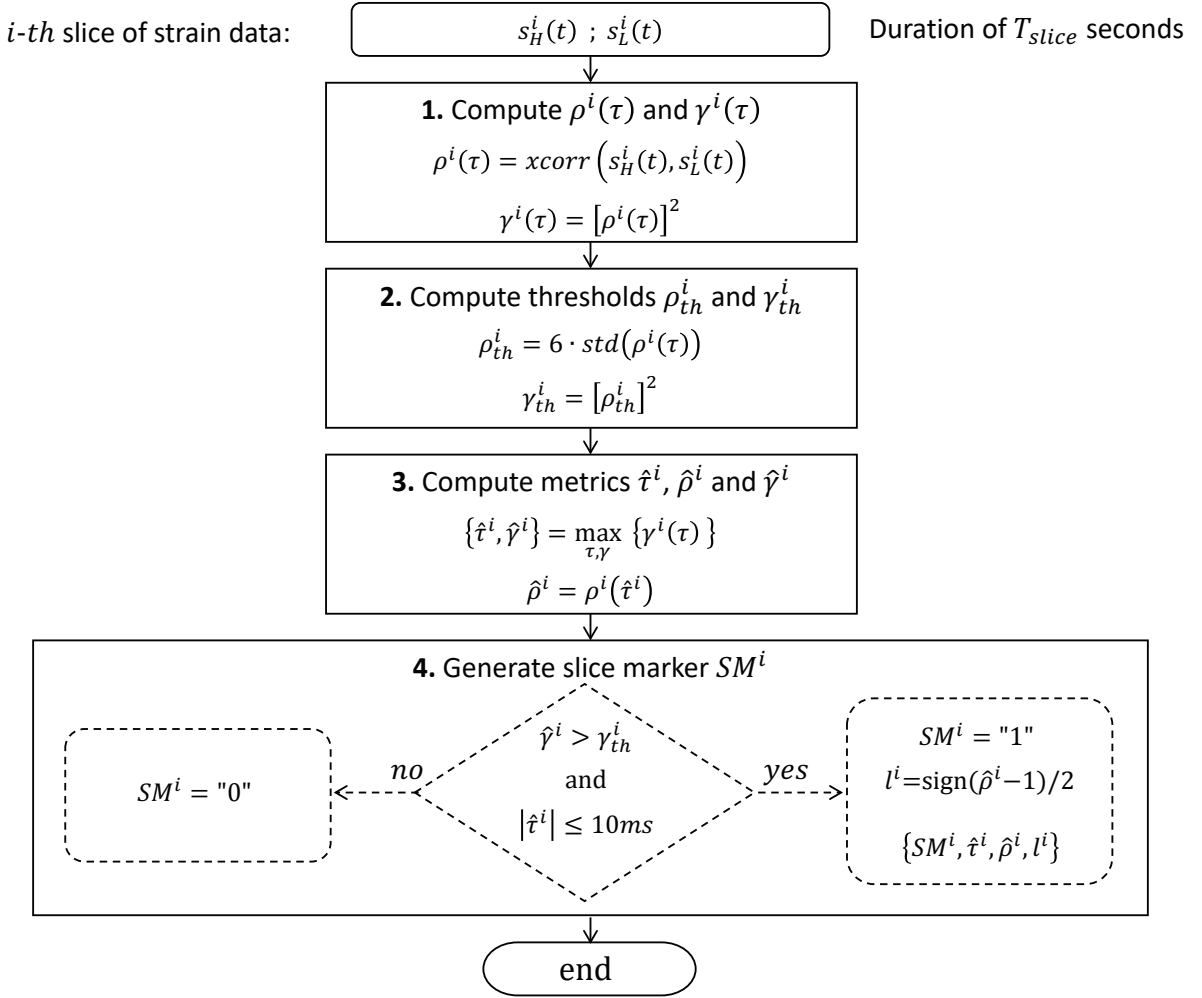}
	\caption{Graphical illustration of the proposed cross-correlation based detector of GW. The input is a short-duration slice of cleaned strain data from the LIGO detectors $s_{H}^{i}(t)$ and $s_{L}^{i}(t)$. The method first computes the cross-correlation $\rho^{i}(\tau)$ and correlation-coefficient $\gamma^{i}(\tau)$ sequences. Then, thresholds are computed for both of them, $\gamma_{th}^{i}$ and $\rho_{th}^{i}$. Subsequently, the maximum correlation-coefficient $\widehat{\gamma}^{i}$ and its corresponding time-shift $\widehat{\tau}^{i}$ and cross-correlation $\widehat{\rho}^{i}$ are computed. Finally, the slice marker $SM^{i}$ is set to one if the maximum correlation-coefficient is greater than the threshold and the time-shift is between $-10$ and $10$ ms, otherwise, the slice marker $SM^{i}$ is set to zero.}
	\label{fig:Fig_Method}
	\end{center}
\end{figure*}

Figure \ref{fig:Fig_Method} shows a graphical illustration of the method.
The input is an $i$-$th$ cleaned strain data slice from LIGO, $s_{H}^{i}(t)$ and $s_{L}^{i}(t)$, of duration $T_{slice}$ seconds, while the output is a slice marker $SM^{i}$ that encodes if a GW signal is present or absent, and the parameters time-shift $\tau^{i}$, cross-correlation value $\rho^{i}$ and the signal inversion factor $l^{i}$ provided that a GW is detected.
The method consists of the following four steps:
\begin{enumerate}
\item \emph{Compute the cross-correlation $\rho^{i}(\tau)$ and correlation-coefficient $\gamma^{i}(\tau)$ sequences.}
\begin{equation}
	\rho^{i}(\tau) = xcorr \left( s_{H}^{i}(t),s_{L}^{i}(t) \right) = \frac{1}{\sqrt{\rho_{H}^{i}(0) \rho_{L}^{i}(0)}} \sum_{j=0}^{N-1} s_{H}^{i}(t_{j}) s_{L}^{i}(t_{j}-\tau)
\end{equation}
\begin{equation}
	\gamma^{i}(\tau) = \left[ \rho^{i}(\tau) \right]^2
\end{equation}
where $\tau=0,\pm t_{1}, \dots ,\pm t_{N-2}$ and $\rho_{H}^{i}(0) = \sum_{j} s_{H}^{i}(t_{j}) s_{H}^{i}(t_{j})$ and $\rho_{L}^{i}(0) = \sum_{j} s_{L}^{i}(t_{j}) s_{L}^{i}(t_{j})$ are the zero-shift auto-correlation of $s_{H}^{i}(t)$ and $s_{L}^{i}(t)$. Note that $\rho^{i}(\tau) \in [-1, 1]$ and $\gamma(\tau) \in [0, 1]$ $\, \forall \, \tau$ and measure the cross-correlation and the correlation-coefficient between $s_{H}^{i}(t)$ and $s_{L}^{i}(t)$ for each time-shift, respectively.
%
\item \emph{Compute the thresholds for the cross-correlation $\rho_{th}^{i}$ and the correlation-coefficient $\gamma_{th}^{i}$}.
\begin{equation}
	\rho_{th}^{i} = 6 \cdot std \left( \rho^{i}(\tau) \right) = 6 \cdot \sqrt{ \frac{ \sum_{k=1}^{M}[\rho^{i}(\tau_{k}) - \overline{\rho^{i}(\tau)} ]^2 }{M-1} }
\end{equation}
\begin{equation}
	\gamma_{th}^{i} = \left[ \rho_{th}^{i} \right]^2
\end{equation}
where $M=2N-1$ is the total number of entries in $\rho(\tau)$, while $\overline{\rho^{i}(\tau)}$ is the average of the cross-correlation sequence $\rho^{i}(\tau$). Note that the cross-correlation threshold $\rho_{th}^{i}$ is six times the standard deviation of the cross-correlation sequence $\rho^{i}(\tau)$.
%
\item \emph{Find the time-shift $\widehat{\tau}^{i}$ and the correlation-coefficient $\widehat{\gamma}^{i}$ value that maximize $\gamma^{i}(\tau)$}.
\begin{equation}
	\lbrace \widehat{\tau}^{i},\widehat{\gamma}^{i} \rbrace = \underset{\tau,\gamma}{\text{max}} \,\, \gamma^{i}(\tau)
\end{equation}
hence, the corresponding cross-correlation value is $\widehat{\rho}^{i}=\rho^{i}(\widehat{\tau}^{i})$.
%
\item \emph{Generate the slice marker $SM^{i}$}. If the maximum correlation-coefficient value $\widehat{\gamma}^{i}$ is greater than the threshold $\gamma_{th}^{i}$ and its corresponding time-shift $\widehat{\tau}^{i}$ is between $-10$ and $10$ ms, then, the slice marker $SM^{i}$ is set to one (this indicates the presence of a GW). Otherwise, the slice marker $SM^{i}$ is set to zero (this indicates the absence of a GW). Finally, if the slice marker indicates that a GW is present, the signal inversion parameter is simply $l^{i} = (sign(\widehat{\rho}^{i})+1)/2$ and the outcome of the method is $\{ SM^{i},\widehat{\tau}^{i},\widehat{\rho}^{i},l^{i} \}$.
\end{enumerate}

\subsection{Data conditioning}
Before applying the proposed cross-correlation based detection method, or any other GW detection method, it is first required to carry out data conditioning (i.e., cleaning) to reduce noise and to remove the common unwanted spectral fluctuations and lines contained in the raw LIGO data.
Because the noise is independent and uncorrelated between the two detectors, data conditioning is performed independently for each LIGO signal.

The first stage of the data conditioning consists of whitening.
Given a long-duration raw strain data segment (typically tens of seconds), for instance from the LIGO Hanford Observatory, $s_{H_{raw}}(t)$, the whitening is computed in the frequency-domain as follows:
\begin{equation}
	s_{H_{white}}(t) = \sum_{k=1}^{N_s} \frac{S_{H_{raw}}(f_k)}{\sqrt{PSD_{H_{raw}}(f_k)}} e^{i 2 \pi t k /N_s}
\end{equation}
where $N_s$ is the number of samples in the segment, $S_{H_{raw}}(f_k)$ is the $N_s$-point discrete Fourier transform and $PSD_{H_{raw}}(f_k)$ is the $N_s$-point two-sided power spectral density (PSD) of the raw data $s_{H_{raw}}(t)$, which can be efficiently and readily computed through the fast Fourier transform (\textit{FFT}) algorithm \cite{proakis1992digital}.
This procedure removes the spectral fluctuations and lines of the raw strain signal leading to a flat power spectrum and its final effect is an enhancement in the signal-to-noise ratio (SNR).
%

The second stage of the data conditioning consists of filtering.
The whitened strain data segment is band-pass filtered in the LIGOs's most sensitive frequency band to remove the low and high-frequency components.
Besides, a portion of the filtered signal at the beginning and the end is discharged to avoid edge effects, this results in a shorter cleaned signal $s_{H}(t)$ for the Hanford observatory and equivalently $s_{L}(t)$ for the Livingston Observatory.

\subsection{Implementation of the method}
\begin{figure*}[t]
	\begin{center}
  	\includegraphics[width=1.00\textwidth]{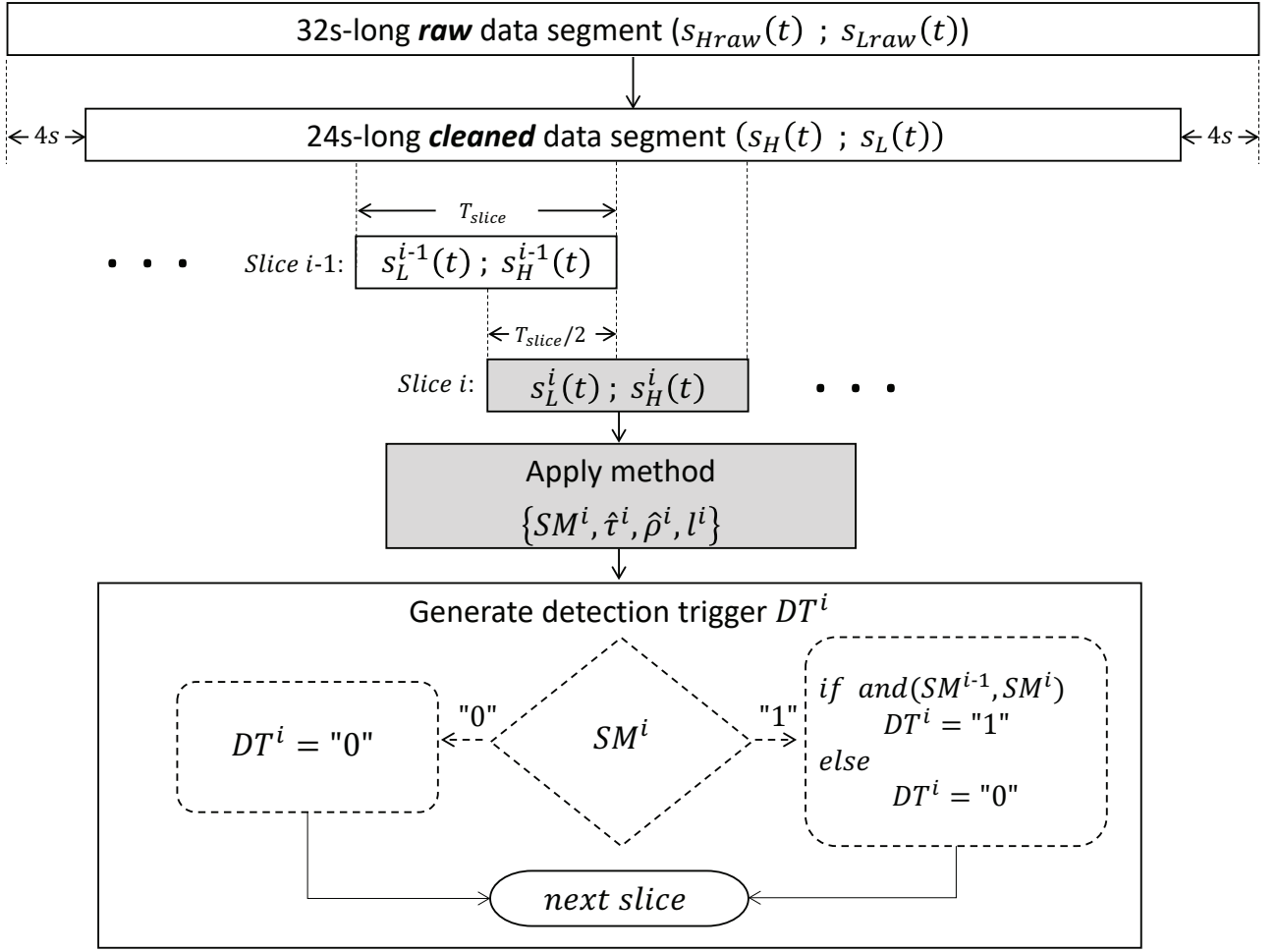}
	\caption{Illustration of the pipeline used to search GW in real LIGO data with the proposed cross-correlation based detector of GW. The input is a 32s-log segment of raw strain data from the LIGO detectors $s_{H_{raw}}(t)$ and $s_{L_{raw}}(t)$. The data conditioning resulted in a 24s-long segment of cleaned strain data. Then, the proposed method is applied to all slices, and a slice marker is generated for each of them. Finally, the detection logic is applied to compute the detection marker.}
	\label{fig:Fig_Pipeline}
	\end{center}
\end{figure*}

Figure \ref{fig:Fig_Pipeline} illustrates the pipeline implemented to search GW using the proposed cross-correlation based detection method.
The input is a 32 s long segment of raw data including the strain from both observatories, $s_{H_{raw}}(t)$ and $s_{L_{raw}}(t)$.
For the whitening, the PSD is computed using the Welch averaged modified periodogram method \cite{welch1967use} with Hanning-windowed epochs of length 4 s and overlap of 2 s.
For the filtering, a zero-phase shift four-order Butterworth-type band-pass filter from 30 to 450 Hz (as this is the LIGOs's most sensitive frequency band) is employed, while the initial and end 4 s of data are discharge.
This results in a 24 s long data segment with cleaned signals $s_{H}(t)$ and $s_{L}(t)$.

Subsequently, the cleaned data segment is divided in consecutive slices of duration $T_{slice}$ seconds with overlap of $T_{slice}/2$.
Slice duration of 0.25, 0.50, 1.00, 2.00, 3.00, 4.00, 6.00 and 8.00 s was considered in this work to assess its effect in detection performance.
Each of these slices is subjected to the proposed cross-correlation based detection algorithm and the slice marker $SM^{i}$ is generated for each slice.

To avoid false positives, we also computed a detection trigger $DT^{i}$ for each slice.
The rationale to compute this trigger is merely that, as two consecutive slices share half of the signal, then if a GW is present in a slice, then, the GW must also be present in one of its neighboring slices.
The detection triggers $DT^{i}$ is computed through the following decision logic.
If two consecutive slice markers $SM^{i-1}$ and $SM^{i}$ were both set to one, then, a GW signal is present and the detection trigger $DT^{i}$ is set to one. Otherwise, no GW signal is present and thus the detection trigger $DT^{i}$ is set to zero.

This pipeline was implemented in MATLAB\textsuperscript{\textregistered} and is freely available at http://www.redtematicaanyog/publicaciones/software. The pipeline is designed to work with data blocks of 4096s-long that can download from the LIGO Open Science Center (LOSC).
%

\section{Application of the method on O1 LIGO data}
\label{sec:application}
%
This section describes the results of the application of the method in search of GW using LIGO data from the first observation run (O1).
The O1 run was carried out from 12 September 2015 0:00 UTC (GPS time 1126051217) until 19 January 2016 16:00 UTC (GPS time 1137254417).
The LIGO's own search of GW using this data culminated in the detection of three GW events generated by binary black hole mergers. Two of these events were confirmed (GW150914 and GW151226) while the other did not achieve the sufficient statistical significance to be accepted as a GW (LVT151012).
Data from O1 was made public on 22 August 2017 to be freely used to study and to perform independent searches of GW \cite{1742-6596-610-1-012021}.
In addition to the confirmed events, O1 data also contains the strain of simulated but realistic GW from compact binary coalescences (CBC) \cite{PhysRevD.95.062002}. 
These simulated GW were injected carefully in both LIGO observatories by adequately moving the arm's mirrors, and this is the reason why they are named hardware injections.
In the present work, O1 data was used to search the three GW events and the CBC hardware injections.
The O1 LIGO data that is freely available to be download consists of 4096s-long data blocks sampled at 4096 Hz, that is, each data block lasts approximately 4.66 hours.
%

\subsection{Searching CBC hardware injections}
\begin{figure*}[t]
	\begin{center}
    \begin{tabular}{ c c }
  	\includegraphics[width=0.50\textwidth]{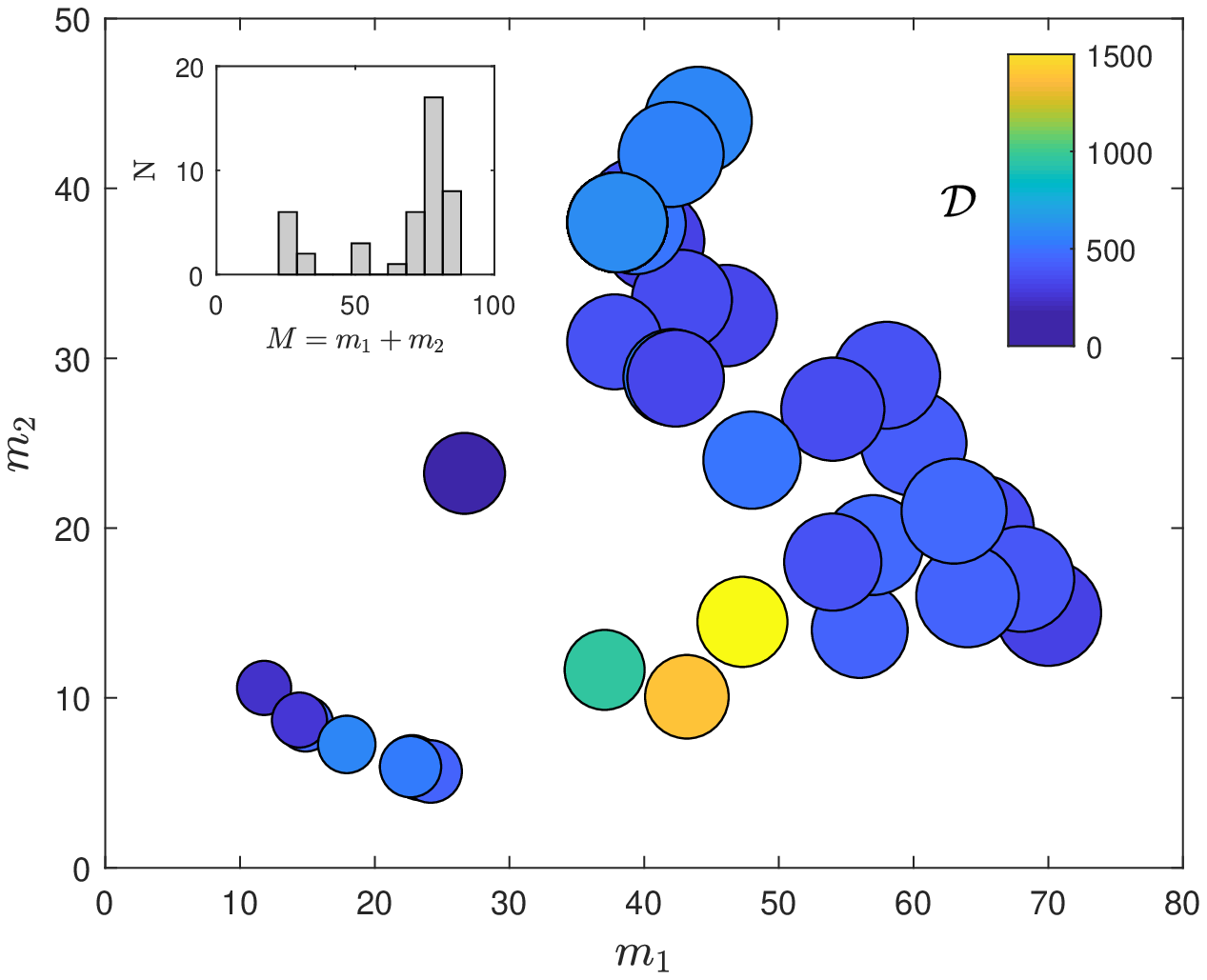}
  	&
  	\includegraphics[width=0.50\textwidth]{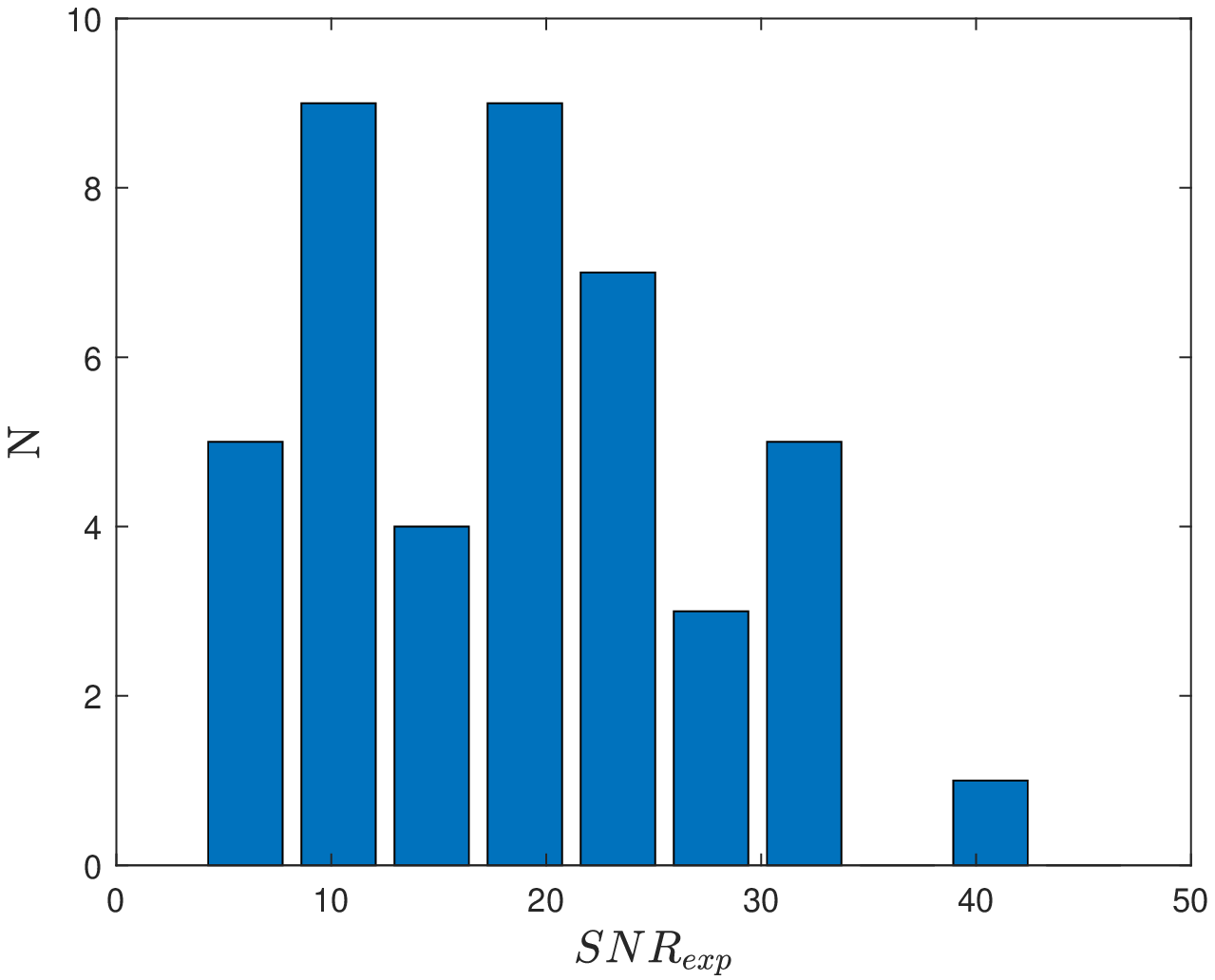}
    \\
    (a) & (b)
	\end{tabular}	
	\caption{(a) Distribution of the CBC masses (in units of sun masses) and distance (in units of Mpc) used to compute the GW that were injected in the two LIGO observatories during the first observation run. The size of the circle encodes the total mass $M=m_1+m_2$, the color encodes the distance and the inner plot depicts the histogram of the total mass $M=m_1+m_2$. (b) Distribution of expected signal-to-noise ratio ($SNR_{exp}$) of the CBC hardware injections.}
	\label{Fig_O1CBCmassdistributions}
	\end{center}
\end{figure*}

All O1 data blocks from the Hanford and Livingston observatories containing a CBC injection were download from the LIGO Open Science Center (LOSC) website at https://doi.org/10.7935/K57P8W9D.
For each injection, the masses of the binary ($m_1$ and $m_2$), the distance of the source ($\mathcal{D}$), the coalescence time ($t_c$) and the expected signal-to-noise ratio ($SNR_{exp}$) are known (see https://losc.ligo.org/o1\_inj/ for a list of the hardware injections and their properties).
In total, there are 62 and 57 CBC hardware injections in the Hanford and Livingston observatory, respectively.
In the present work, we only consider data blocks that contain the complete strain data, that passed all quality checks, and that contain simultaneous injection in the two observatories, this resulted in a total of 42 CBC hardware injections that were used in the rest of this study.
Figure \ref{Fig_O1CBCmassdistributions}a shows the distribution of the masses ($m_1$ versus $m_2$) with its corresponding distance distance ($\mathcal{D}$).
For all injections, the distance of the CBC system ranged from 154.12 to 1594.67 Mpc.
The upper left part of the figure shows the histogram of total mass $M=m_1+m_2$ which ranged from 22.38 to 88.00 solar masses.
Figure \ref{Fig_O1CBCmassdistributions}b shows the distribution of $SNR_{exp}$ for these injections. The $SNR_{exp}$ ranges from 3.80 to 42.56.

\begin{figure*}[t]
	\begin{center}
    \begin{tabular}{ c c }
  	\includegraphics[width=0.50\textwidth]{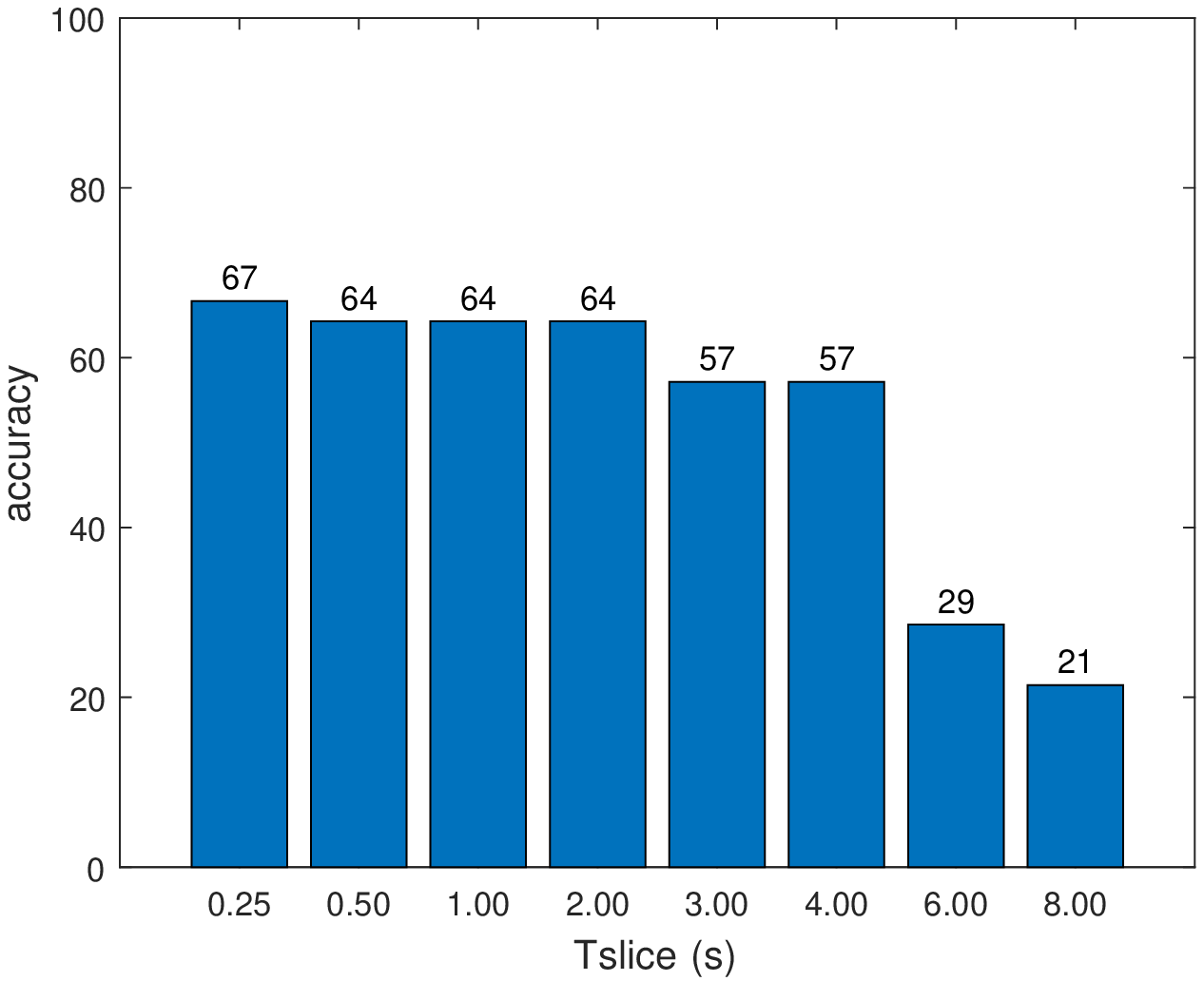}
  	&
  	\includegraphics[width=0.475\textwidth]{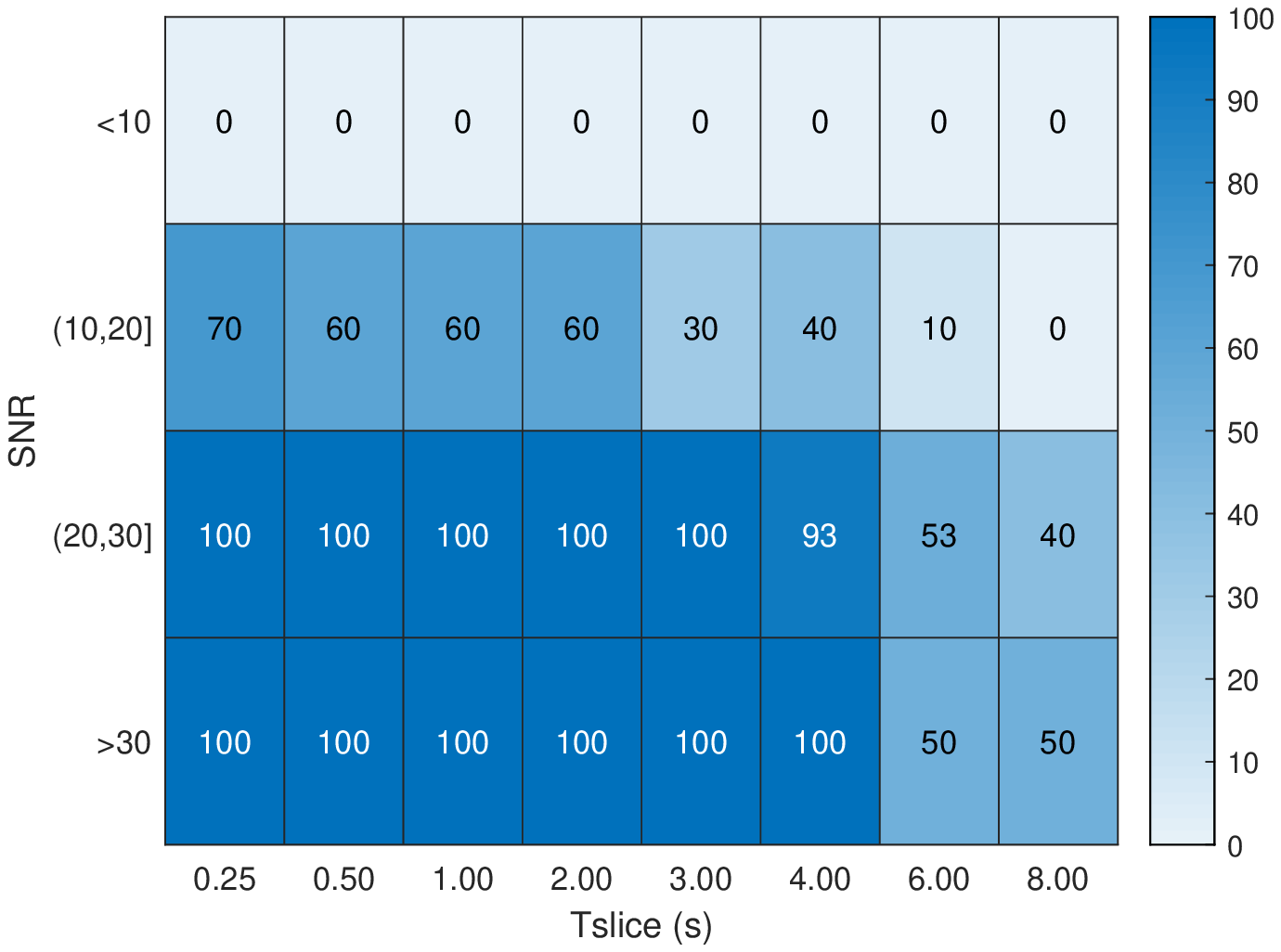}
    \\
    (a) & (b)
	\end{tabular}
	\caption{(a) Accuracy in the detection of the hardware injections achieved with the proposed cross-correlation method for several values of slice duration $T_{slice}$ where the search is carried out. (b) Detection accuracy for different ranges of SNR for each $T_{slice}$. These results show; first, the detection accuracy is better for small slice duration, and second, the greater the SNR, the better the detection accuracy.}
	\label{Fig_ACCvsTslice}
	\end{center}
\end{figure*}

To assess performance, the percentage of correctly detected injections was used to measure detection accuracy. 
This metric was computed as follows:
\begin{equation}
	accuracy = 100 \times \frac{Number \, of \, detections}{Number \, of \, injections} \, ,
\end{equation}
where $Number \, of \, injections = 42$.
Figure \ref{Fig_ACCvsTslice}a shows the detection accuracy achieved with the proposed cross-correlation method for the several values of slice duration $T_{slice}$, i.e., 0.25, 0.50, 1.00, 2.00, 3.00, 4.00, 6.00 and 8.00 s.
These results show that the highest detection accuracies are achieved for small values of $T_{slice}$ while detection accuracy is reduced for high values of it.
This is because all injected CBC gravitational waves are of short-duration (the duration of the GW signal was computed for all injections according to Eq. 3.5 of Ref. \cite{PhysRevD.85.122006}, and it was on average 0.3899$\pm$0.3235 s with minimum of 0.1658 and maximum of 1.0692 s), therefore, for small values of $T_{slice}$ the slice of strain data employed to perform the search contains mainly GW signal and thus the correlation-coefficient is high. On the contrary, long values of $T_{slice}$ contain much noise, and thus the correlation-coefficient values are decreased.

To examine the influence of the SNR of the injected GW signals in the detection accuracy, figure \ref{Fig_ACCvsTslice}b presents the detection accuracy for the several slice duration $T_{slice}$ and for different ranges of SNR.
These results show that the best performance is achieved for high values of SNR while detection accuracy degrades for low values of it.
Indeed, it is observed that for SNR in the ranges of $(20,30]$ and $>30$ all injections were detected except for those where the search is performed with slices of duration $6.00$ and $8.00$ s.
On the contrary, for the lower range of SNR (i.e., $SNR<10$) no injections were detected regardless of the duration of the slice $T_{slice}$.

\begin{figure*}[t]
	\begin{center}
    \begin{tabular}{ c }
  	\includegraphics[width=1.0\textwidth]{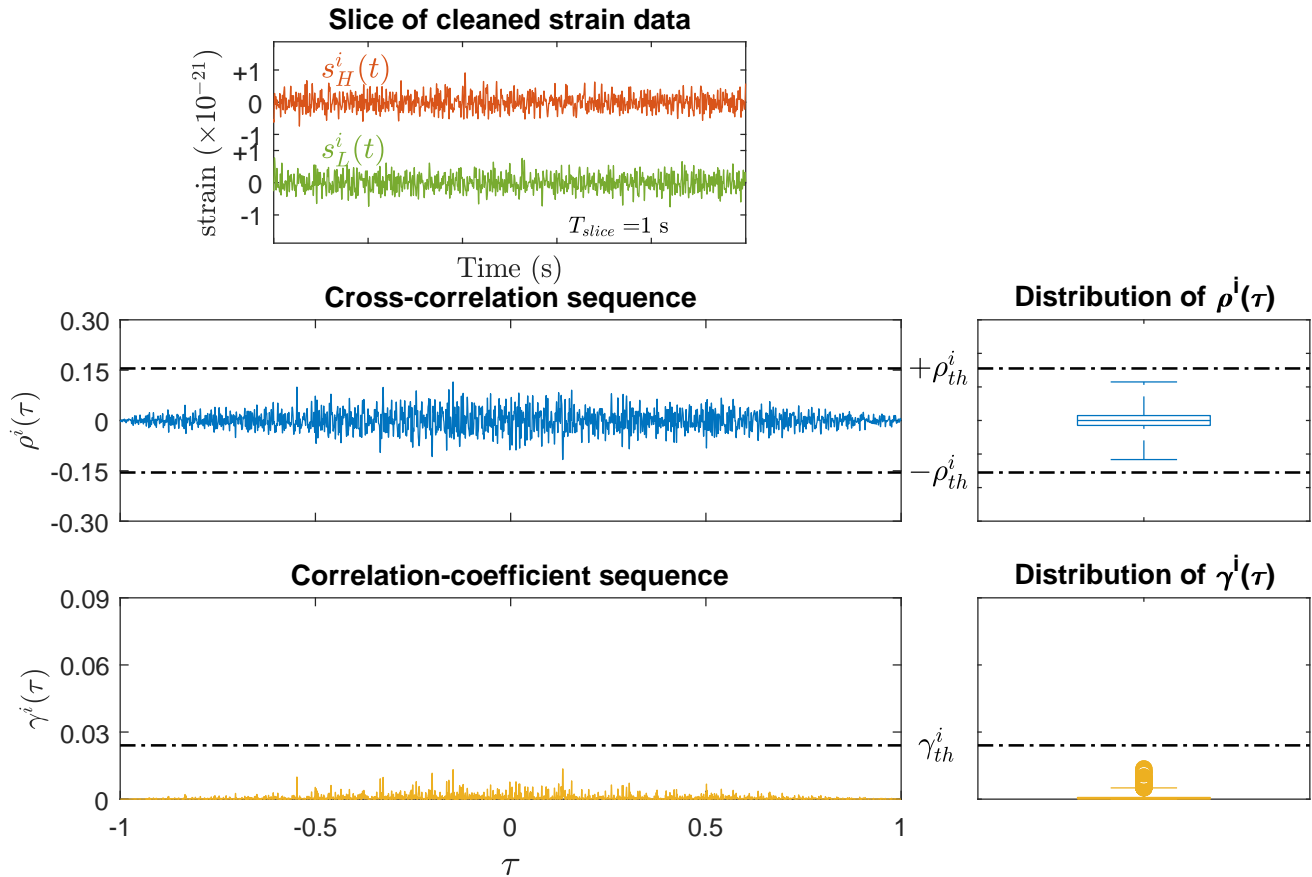}
    \end{tabular}
	\caption{
Illustration of the method for a slice of LIGO data of duration $T_{slice}=$1 s where no GW signal is present.
Upper panel: slice of cleaned strain data, $s_{H}^{i}(t)$ and $s_{L}^{i}(t)$, where no transients are observed.
Central panel: cross-correlation sequence $\rho^{i}(\tau)$ and its corresponding threshold $\rho_{th}^{i}$; No positive/negative peak in $\rho_{th}^{i}$ is greater/lower than the threshold; The distribution of $\rho^{i}(\tau)$ shows no values that are greater or lower than $\rho_{th}^{i}$.
Lower panel: correlation-coefficient sequence $\gamma^{i}(\tau)$ and its corresponding threshold $\gamma_{th}^{i}$; No peak greater than $\gamma_{th}^{i}$ is observed in $\gamma^{i}(\tau)$; The distribution of $\gamma^{i}(\tau)$ shows no values that are greater than $\gamma_{th}^{i}$.
As no correlation-coefficient value is greater than the threshold, then no GW is detected and the slice marker $SM^i$ is set to zero.
}
	\label{Fig_MethodXcorrelCase1}
	\end{center}
\end{figure*}

\begin{figure*}[t]
	\begin{center}
    \begin{tabular}{ c }
    \includegraphics[width=1.0\textwidth]{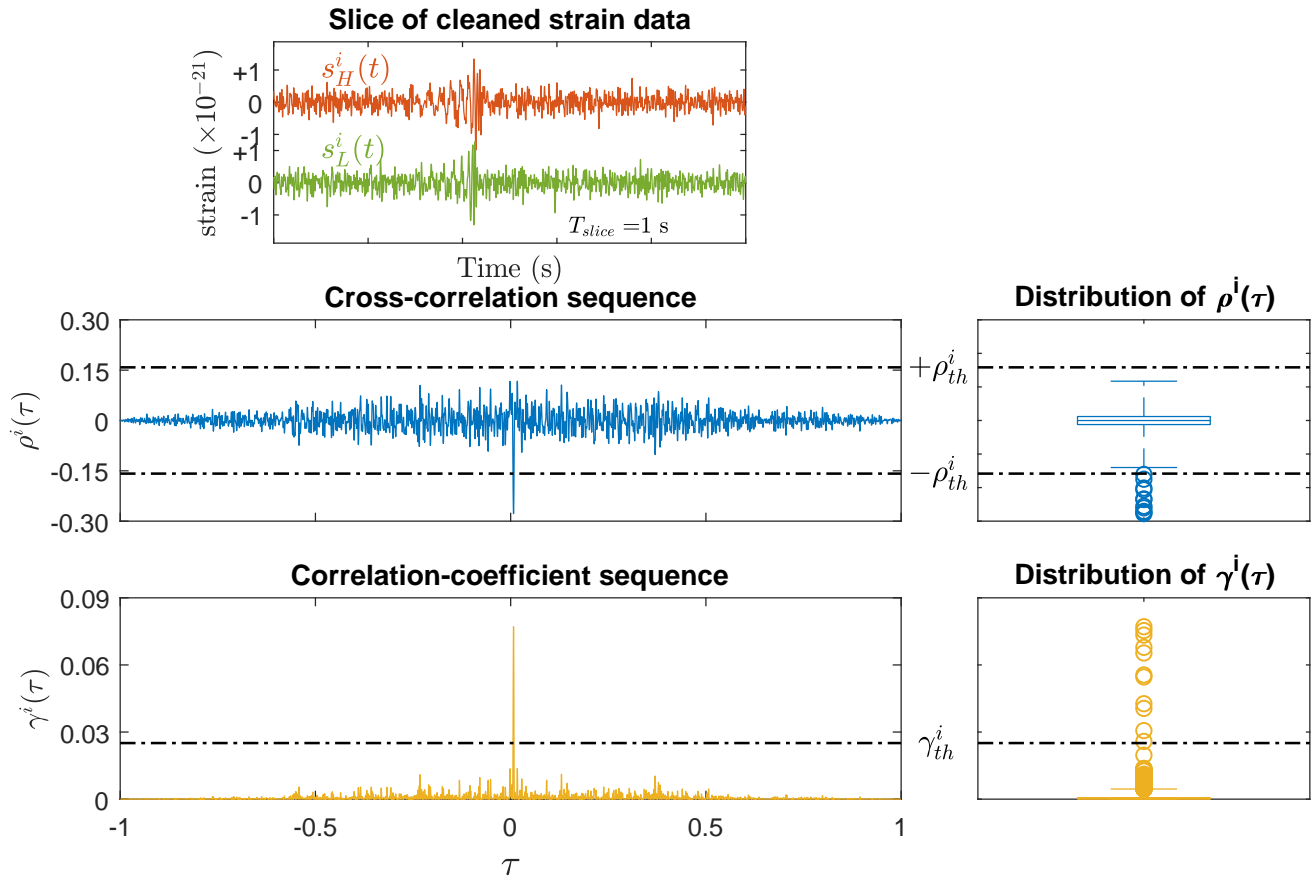}
    \end{tabular}
	\caption{
Illustration of the method for a slice of LIGO data of duration $T_{slice}=$1 s where a GW signal is present. 
Upper panel: slice of cleaned strain data, $s_{H}^{i}(t)$ and $s_{L}^{i}(t)$, where a seemingly simultaneous transitory activity is observed in the two signals.
Central panel: cross-correlation sequence $\rho^{i}(\tau)$ and its corresponding threshold $\rho_{th}^{i}$; Notice that a negative peak in $\rho^{i}(\tau)$ is lower than $-\rho_{th}^{i}$; The distribution of $\rho^{i}(\tau)$ also shows values that are lower than $-\rho_{th}^{i}$.
Lower panel: correlation-coefficient sequence $\gamma^{i}(\tau)$ and its corresponding threshold $\gamma_{th}^{i}$; A peak greater than $\gamma_{th}^{i}$ is observed in $\gamma^{i}(\tau)$; The distribution of $\gamma^{i}(\tau)$ also shows values that are greater than $\gamma_{th}^{i}$.
As the maximum correlation-coefficient value is greater than the threshold and its corresponding time-shift is between $\pm 10$ ms, then a GW is present and the slice marker $SM^i$ is set to one.
}
	\label{Fig_MethodXcorrelCase2}
	\end{center}
\end{figure*}

To illustrate the idea behind the method, figures \ref{Fig_MethodXcorrelCase1} and \ref{Fig_MethodXcorrelCase2} show two slices of cleaned strain data of duration 1 s with their corresponding cross-correlation and correlation-coefficient sequences, the distribution for both of them and their thresholds.
Figure \ref{Fig_MethodXcorrelCase1} shows these results for a slice of strain data recorded from GPS time 1127789364 where no GW signal, neither hardware injection nor a real event, is present.
The cleaned strain data is presented in the upper panel, shows no evidence of transitory activity and only noisy signals are observed.
The cross-correlation sequence $\rho^{i}(\tau)$ is presented in the central panel. The minimum value is -0.1165, the maximum value is 0.1149, and the standard deviation is 0.0258. These results show no value of $\rho^{i}(\tau)$ above or below the corresponding threshold $\rho_{th}^{i}$; this is also observed in the distribution presented on the right where no single value is above or below the threshold.
The correlation-coefficient sequence $\gamma^{i}(\tau)$ is presented in the lower panel and has a maximum value of 0.0136 and a standard deviation of 0.0012. Hence, the threshold $\gamma_{th}^{i}$ is 0.0240, this shows that no value of $\gamma^{i}(\tau)$ is above $\gamma_{th}^{i}$, which is also observed in the distribution presented on the right.
In this case, no correlation-coefficient value is greater than the correlation-coefficient threshold; therefore, no GW is detected. 

Figure \ref{Fig_MethodXcorrelCase2}, on the other hand, shows the results for a slice of strain data recorded from GPS time 1127789374 where an injected GW signal is present.
The cleaned strain data presented in the upper panel shows that the two strain signals have a seemingly simultaneous transient activity at about the middle part of the slice.
The cross-correlation sequence $\rho^{i}(\tau)$ is presented in the central panel and has a minimum value of -0.2776, a maximum value of 0.1171 and a standard deviation of 0.0264. A negative peak in $\rho^{i}(\tau)$ is observed which is lower than the negative threshold $-\rho_{th}^{i}$, moreover, this peak matches with the transient observed in the cleaned strain data. The distribution presented on the right also shows values of $\rho^{i}(\tau)$ that are below the negative threshold.
The correlation-coefficient sequence $\gamma^{i}(\tau)$  is presented in the lower panel and has a maximum value of 0.0771 and a standard deviation of 0.0024. Hence, the threshold $\gamma_{th}^{i}$ is 0.0251. 
It is easily observed that the maximum value of $\gamma^{i}(\tau)$ is greater than $\gamma_{th}^{i}$, which is also observed in the distribution presented on the right.
For this case, the corresponding time-shift is $\widehat{\tau}^{i}=7.3242$ ms which is within the range of $\pm 10$ ms; consequently, a GW is present. 

%


\subsection{Searching the events GW150914, GW151226 and LVT151012}
The method was also used to search the events GW150914, GW151226 and LVT151012.
To this end, the three 4096s-long data blocks containing these events were downloaded from the LOCS \cite{1742-6596-610-1-012021} website and subjected to the search pipeline using the same slice durations $T_{slice}$ of 0.25, 0.50, 1.00, 2.00, 3.00, 4.00, 6.00 and 8.00 s.

For GW150914, table \ref{Tab_ResultsGW150914} presents the detection results obtained for the different values of the slice duration.
Detection was achieved successfully for slice duration of 0.25, 0.50, 1.00, 2.00 and 3.00 s, while no detection was achieved for slice duration of 4.00, 6.00 and 8.00 s.
Note that GW150914 had a duration of 0.2 s while swept upwards from 35 up to 250 Hz \cite{PhysRevLett.116.061102,PhysRevX.6.041015}, that is, it was a short-duration event which is the reason why it is not detected with long duration slices.
All slices with successful detection presented a time-shift $\widehat{\tau}^{i}$ of around 7.32 ms, which is an inter-site propagation time lower than the bound of 10 ms.
Regarding the cross-correlation value $\widehat{\rho}^{i}$, it is observed that the larger the duration of the slice the lower the correlation between the two LIGO signals; this is explained again by the fact that GW150914 is a short-duration event and thus for long-duration slices, the strain data contains more uncorrelated noise than correlated GW signal.
Finally, in all the cases where detection was achieved, the signal inversion parameter was 1 which indicates that the GW signal arrived first to the L1 observatory and then to the H1 observatory. This agrees with the results reported in the seminal publications of the LIGO team for this event \cite{PhysRevLett.116.061102}. 
\begin{table}[h]
\caption{Detection marker, time-shift, cross-correlation value and inversion parameter obtained for the event GW150914 with the different values of slice duration. Detection was achieved for the shortest slice duration of 0.25, 0.50, 1.00, 2.00 and 3.00 s but no for the longest slice duration of 4.00, 6.00 and 8.00 s. For slices where detection was achieved, the time-shift $\widehat{\tau}^{i}$ is around 7.32 ms, the cross-correlation approaches to zero as the slide duration increases and the signal inversion parameter is -1 which indicates that GW signal arrived first to L1 and then to H1.
}
\label{Tab_ResultsGW150914}
\begin{center}
\begin{tabular}{ccccc}
\hline
$\,T_{slice}\,$ & $DM^{i}$ & $\widehat{\tau}^{i}$ & $\widehat{\rho}^{i}$ & \,\,\, $l^{i}$ \,\,\, \\
\hline
\hline
0.25        & 1        & 7.32                 & -0.61                & 1      \\
0.50        & 1        & 7.32                 & -0.44                & 1      \\
1.00        & 1        & 7.32                 & -0.25                & 1      \\
2.00        & 1        & 7.32                 & -0.14                & 1      \\
3.00        & 1        & 7.08                 & -0.11                & 1      \\
4.00        & 0        & -                    & -                    & -       \\
6.00        & 0        & -                    & -                    & -       \\
8.00        & 0        & -                    & -                    & -       \\
\hline
\end{tabular}
\end{center}
\end{table}

For GW151226 and LVT151012, no GW signal was detected irrespective of the slice duration.
Note however that the LIGO's matched filter based search was able to detect these events \cite{PhysRevLett.116.241103,PhysRevX.6.041015} with a SNR of 13 and 9.7, respectively.
The low SNR indicates that these events are weaker than the observation noise and that they do not stand above the noise even after whitening and band-pass filtering.
This is precisely the reason why it was not possible to detect them.
These no detection results agree with the low detection accuracy for low SNR achieved in the previous search of CBC hardware injections.
\section{Discussion}
\label{sec:discussion}

%
The first astrophysical observations announced by the LIGO and VIRGO scientific collaboration \cite{0034-4885-72-7-076901,0264-9381-32-7-074001,AdvancedVirgo2015} demonstrated the existence of the gravitational waves (GW) signals predicted in the general relativity theory \cite{Einstein15a,Einstein15b}, and showed the existence of binary black holes (BBH) \cite{PhysRevLett.116.061102,PhysRevLett.116.241103,PhysRevLett.118.221101,2041-8205-851-2-L35,PhysRevLett.119.141101} and binary neutron stars (BNS) \cite{PhysRevLett.119.161101}.
The essential components in these and upcoming observations are the theoretical and numerical solutions of the Einstein's equations, the earth-based network of detectors used to measure the very tiny strain signals induced by GW generated by distant astrophysical sources, and the data analysis and processing methods devoted to detect and extract GW signals contained in the noisy observations recorded from the observatories \cite{Antelis2017}.
This work focuses precisely on the last component and proposes a method for the detection of GW signals that relies on cross-correlation and phenomenological information.
This method is used to search injected and real GW signals contained in LIGO's first observation run (O1).

%
The proposed method computes the similarity across time-shifts, i.e., time-domain cross-correlation, between short-duration cleaned strain signals from the LIGO observatories and decided if a GW signal is present based on phenomenological information.
Concretely, the idea is that if the maximum correlation-coefficient between the LIGO signals is superior to a significant threshold and its corresponding time-shift is between $-10$ and $10$ ms; then a GW signal is present in the data under scrutiny.
Unlike other methods as the matched filter where the search is applied independently for each detector and a detection trigger is generated by coincidence detections across detectors \cite{PhysRevD.85.122006}, this method processes simultaneously the strain data from the detectors to generate a detection trigger.
%
%
Hence, the advantage is that it employs simultaneously the data from the two detectors while no assumptions about the waveform of the GW signals are needed, i.e., this is a template-free method.
Because of this, the method can be used to search any type of GW irrespective of its astrophysical source, location and waveform.
However, since the method is based entirely on the time-domain cross-correlation, it is required a large signal-to-noise ratio.

%
It is important to mention three aspects of the method.
First, the cross-correlation is computed across all available time-shifts, i.e., up to the length of the signals. However, it is also possible to restrict its calculation to time lags between $-10$ and $10$ ms and then to decide if a GW is present by examining if the maximum correlation-coefficient is greater than the corresponding threshold. Although this is similar to our method, we preferred to use all time lags as it allows to obtain more data and therefore more confidence in the decisions.
Second, the method can be extended easily to multiple detectors, i.e., the strain data from VIRGO \cite{AdvancedVirgo2015}, GEO \cite{GEO2014}, KAGRA \cite{KAGRA} and other future detectors as LIGO-India can also be incorporated. This can be done by simply computing cross-correlations between the strain data across all available detectors and then to decide if a simultaneous transient is presently provided that each time-shift is between the corresponding inter-observatory propagation time. This would be an interesting search tool for the second observation run (O2) and for the upcoming third observation run (O3), which comprises a network of three detectors including the LIGO Hanford and Livingston observatories along with the VIRGO observatory.
Third, the method does not provide any information about the properties and the location of the astrophysical source responsible for the detected GW; however, this is critical and necessary for GW astronomy. One way to address this issue is to use the strain data where the GW is detected as input to post-processing steps to determine more physical information. In this line, recent works have proposed the use of machine learning algorithms to estimate astrophysical source parameters based on the observed strain data \cite{Carrillo2016,GEORGE201864}.

%
The method was used to search injected and real GW signals contained in O1.
On the one hand, the search of injected GW signals showed that the highest detection accuracies were achieved for short duration strain data (see figure \ref{Fig_ACCvsTslice}a) and for high values of SNR (see figure \ref{Fig_ACCvsTslice}b). The higher performance for small duration data slices was due to the short duration of all injected GW data and not a particular property of the method, while the higher performance for high values of SNR was simply because the cross-correlation increases to the extent that the amplitude of GW signal also increases. 
On the other hand, the search for the real GW events resulted in the detection of GW150914 but not of GW151226 and LVT151012. 
GW150914 was detected with the shortest data slices of 0.25, 05, 1.0, 2.0 and 3.0 s  (see Table \ref{Tab_ResultsGW150914}), which is understandable as this event had a duration of 0.2 s \cite{PhysRevLett.116.061102,PhysRevX.6.041015}. All detections presented a time-shift of around 7.32 ms, which is an inter-site propagation time lower than the bound of 10 ms; moreover, the inversion parameter indicated that the GW signal arrived first to the Livingston observatory and then to the Hanford observatory. This agrees with the results reported in the seminal publications of the LIGO team for this event \cite{PhysRevLett.116.061102}.
GW151226 and LVT151012 were not detected irrespective of the slice duration. Note that the LIGO's search pipeline detected these events with a SNR of 13 and 9.7, respectively \cite{PhysRevLett.116.241103,PhysRevX.6.041015}. The low SNR indicates that these two signals are much weaker than the noise which is precisely the reason why the method was not able to detect them. This agrees with the results presented in search of injected GW where reduced detection accuracy was achieved for GW signals with low SNR.

\section{Conclusion}
\label{sec:conclusion}
%
This work presents an independent method to search GW signals using LIGO data.
The method relies on cross-correlation and phenomenological information, specifically, GW is detected if the strain data where the search is performed presents a significant correlation-coefficient with a time-lag that is no greater than the expected inter-site propagation time.
Hence, it is a basic template-free framework that can be used to search GW irrespective of the GW signatures and the properties of the astrophysical source; however, large signal-to-noise ratios are required to be able to find significantly correlated signals embedded in the noise.
The method is not intended to replace the complex and outstanding GW detection, parameter estimation and source localization pipelines carried out by LIGO, but it is a basic computational tool to perform and independent and rapid search of GW irrespective of its waveform, astrophysical origin and sky location.

%
\section*{Acknowledgements}
This research has made use of data, software and/or web tools obtained from the LIGO Open Science Center (https://losc.ligo.org), a service of LIGO Laboratory, the LIGO Scientific Collaboration and the Virgo Collaboration. The U.S. National Science Foundation funds LIGO. The French Centre National de Recherche Scientifique (CNRS), the Italian Istituto Nazionale della Fisica Nucleare (INFN) and the Dutch Nikhef, with contributions by Polish and Hungarian institute funds VIRGO.
The authors would like to thank the support of the CONACyT grant No. 294625 and the CONACyT-AEM grants No. 262847. CM thanks the support of PROSNI-UDG 2018 and PRODEP UDG-CA-813.

%
\bibliography{references}

%
\end{document}